\documentstyle[12pt]{article}

\newcommand{\Om}{\Omega}
\newcommand{\om}{\omega}
\newcommand{\ep}{\varepsilon}
\newcommand{\gm}{\gamma}
\newcommand{\dgr}{\dagger}
\newcommand{\ov}{\overline}
\newcommand{\bt}{\beta}
\newcommand{\al}{\alpha}
\newcommand{\Lb}{\Lambda}
\newcommand{\ra}{\rightarrow}
\newcommand{\kp}{\kappa}
\newcommand{\prt}{\partial}
\newcommand{\be}{\begin{equation}}
\newcommand{\ee}{\end{equation}}

\begin{document}

\begin{center}
DO WE UNDERSTAND WHAT IS DECONFINEMENT?  \\ [5mm]

V.I. Yukalov$^1$ and E.P. Yukalova$^2$ \\ [3mm]

{\it $^1$ Bogolubov Laboratory of Theoretical Physics \\
Joint institute for Nuclear Research, Dubna 141980, Russia \\ [3mm]

$^2$ Laboratory of Informational Technologies \\
Joint institute for Nuclear Research, Dubna 141980, Russia}

\end{center}

\begin{abstract}

An overview is given of different approaches to describing the process
of deconfinement in quantum chromodynamics. The analysis of the known 
approaches demonstrates that the detailed picture of how deconfinement 
really occurs has not yet been understood. Therefore, one has to be 
rather cautious when interpreting experimental signals as attributed to 
deconfinement.

\end{abstract}

\vskip 1cm

{\bf Key-words}: deconfinement, phase transitions, crossovers, 
quark-hadron matter.

\vskip 2cm

\section{Introduction}

When talking about deconfinement in quantum chromodynamics, one should 
distinguish two things: deconfinement as a {\it general phenomenon} and 
deconfinement as a {\it concrete process}. What is, in general, the phenomenon
of deconfinement is well known - this is the transformation of hadron 
matter into quark-gluon plasma occurring with increasing temperature or 
density of matter (see reviews [1--5]). But how this process occurs in 
reality? The answer to the question is crucially important for the correct 
interpretation of experimental signals, that are attributed to 
deconfinement, such as the suppression of charmonium production ($J/\psi$ 
suppression [5,6]), the enhancement of strangeness production [7,8], the 
enhancement of dilepton production [7,8] or other signals.

The process of deconfinement can be considered from two points of view, 
{\it stationary} and {\it nonstationary}. In the stationary picture, one 
considers the transformation of {\it equilibrium infinite} hadron matter 
into quark-gluon plasma, when rising temperature or density. In the 
nonstationary picture, one takes into account the peculiarities related to 
heavy ion reactions, which involves the consideration of deconfinement in 
{\it nonequilibrium finite} objects [9]. As is evident, before going to the 
complications of the nonstationary picture, it is necessary to have a more 
or less complete understanding of what happens in the stationary picture, 
which the nonstationary one is based on. In what follows, we shall analyze 
only the stationary case and will show that even this has not yet been 
completely understood.

\section{Numerical Lattice Simulations}

Lattice simulations of high-temperature QCD provide nonperturbative 
theoretical insights into the phenomenology of the transition from hadronic 
matter to the quark-gluon plasma. One of the basic goals of lattice QCD 
calculations at finite $T$ is to provide quantitative results for the
deconfinement transition temperature. Up to now, this goal has been 
achieved only in the pure gauge sector [10]. The value of the critical 
temperature for the deconfining phase transition in a $SU(3)$ pure gauge 
theory (without quarks) is known with small errors of the order of $3\%$.
This temperature $T_c\approx 270$ MeV, and deconfinement is a real phase 
transition, either of first order or a rather sharp phase transition of 
second order [10].

Unlike in the pure gluodynamics, the transition temperature for $SU(3)$ 
chromodynamics with finite quark masses is not well defined. Even at 
vanishing baryon number density, there is no yet a satisfactory 
understanding of the critical behaviour in QCD [10]. In zero-density
chromodynamics at physical values of quarks, 
deconfinement is rather a rapid crossover than a pure phase transition 
[10,11]. The crossover temperature can conditionally be defined as that 
corresponding to a maximum of some thermodynamic characteristics [12], such 
as a susceptibility [10,11]. Different lattice calculations for the 
zero-density QCD give the crossover temperatures in the interval 
$T_c\approx (140-190)$ MeV [10,11].

Finite baryon density calculations in QCD are affected by the so-called 
sign problem, when the fermion determinant becomes complex for nonzero 
values of the chemical potential and the partition function fails to be 
positive [10,13-15], which prohibits the use of the conventional numerical 
algorithms. Because numerical simulations of QCD at finite baryon density 
are plagued by the principal technical difficulties, the present 
understanding of deconfinement is not satisfactory. However, the available 
data show no signals of phase transition [13--15]. The present situation is 
rather pessimistic - it seems that there is no reliable hope to get 
important improvements in the knowledge of QCD at finite density from 
lattice simulations [14].

\section{Pure Phase Models}

Because of the principal difficulties and uncertainty in the finite-density 
lattice simulations, several phenomenological models of deconfinement have 
been suggested. The most often employed are the pure phase models, when 
hadron matter and quark-gluon plasma are treated as different pure 
thermodynamic phases. Each phase is supposed to possess its own 
thermodynamic potential.

It is convenient to work with the grand potential $\Om=-PV$, which is a 
function of temperature $T$, volume $V$, and chemical potentials $\mu_i$ of 
different particles, each particle sort being enumerated by the index 
$i=1,2,\ldots$.  Each type of particles is characterized by a set of 
quantum numbers, such as the baryon number $B_i$, strangeness $S_i$, and 
others. The related baryon density $n_B$ and strangeness density $n_S$ are
\be
n_B =\sum_i B_i\; \rho_i \; \qquad n_S =\sum_i S_i\; \rho_i \; ,
\ee
where $\rho_i$ is the density of particles of type $i$. Then the chemical 
potential $\mu_i$ can be expressed as
\be
\mu_i =\mu_B B_i + \mu_S S_i \; ,
\ee
with the baryon potential $\mu_B$ and strangeness potential $\mu_S$. 
Therefore, the grand potential can be considered as a function 
$\Om=\Om(T,V,\mu_B,\mu_S)$. Consequently, the pressure is a function 
$P=P(T,\mu_B,\mu_S)$. The baryon and strangeness densities (1) can be 
written as the derivatives
\be
n_B=\frac{\prt P}{\prt \mu_B} \; , \qquad n_S =
\frac{\prt P}{\prt \mu_S} \; .
\ee
In what follows, we shall consider, for simplicity, the case with fixed 
$\mu_S$ and will analyze the dependence of pressure on temperature and $\mu_B$,
writing $P=P(T,\mu_B)$. Respectively, the baryon density $n_B=n_B(T,\mu_B)$.

In pure-phase models of deconfinement, one divides all particles into two 
groups, one group consisting of hadrons and another group consisting of 
quarks and gluons. The first group is assumed to form the hadron phase and 
the second one, the quark-gluon phase. The corresponding  pressures of pure 
phases, $P_h$ and $P_p$, are calculated in different approximations [1--5]. 
As a result, the baryon densities of pure hadron and pure plasma phases are 
different,
\be
n_{Bh} \equiv \sum_i B_{ih}\; \rho_{ih} =\frac{\prt P_h}{\prt\mu_B} \; ,
\qquad n_{Bp} \equiv \sum_i B_{ip}\; \rho_{ip}=
\frac{\prt P_p}{\prt\mu_B} \; ,
\ee
where the summations include, respectively, either only hadrons (and 
antihadrons) or only  quarks, antiquarks, and gluons. The 
deconfinement temperature is given by the equality
\be
P_h(T_c,\mu_B) = P_p(T_c,\mu_B)
\ee
yielding a uniquely defined transition line $T_c=T_c(\mu_B)$. However, the 
transition temperature as a function of baryon density is not uniquely defined.
This is because there are two different baryon densities (4), which also 
are different at the transition temperature,
\be
n_h \equiv n_{Bn}(T_c,\mu_M) \; , \qquad
n_p\equiv n_{Bp}(T_c,\mu_B) \; .
\ee
These densities are different at $T_c$ since at this point the pressures
of hadron phase and of plasma phase intersect. For the deconfinement 
transition, one has $n_h>n_p$. Treating the relations (6) as the equations 
for the baryon potential, one gets two different potentials $\mu_h\equiv 
\mu_B(T_c,n_h)$ and $\mu_p\equiv \mu_B(T_c,n_p)$. Substituting these into 
the dependence $T_c(\mu_B)$, one obtains two lines
\be
T_h(n_h) \equiv T_c(\mu_h) \; , \qquad T_p(n_p) \equiv T_c(\mu_p) \; .
\ee
These two lines environ the region on the plane $T_c-n_B$ where the 
hadron and plasma phases coexist. This situation is completely analogous 
to the standard case of first-order phase transition, when coexisting 
phases are treated in different approximations or when the pressure as a 
function of density contains an instability interval [16,17]. In the 
considered case, the baryon potential of the hadron-plasma mixture is
\be
\mu_B(T_c,n_B) =x_h\; \mu_h + x_p\; \mu_p \; ,
\ee
where the phase concentrations are defined by the equations
\be
x_h n_h + x_p n_p = n_B \; , \qquad x_h + x_p =1 \; ,
\ee
which yield
\be
x_h =\frac{n_B-n_p}{n_h-n_p} \; , \qquad 
x_p = \frac{n_h-n_B}{n_h-n_p} \; .
\ee
Let us stress that the coexisting phases, characterized by the linear 
combination (8), are both macroscopic. Such a mixture of {\it macroscopic 
phases} is called the {\it Gibbs mixture}, in order to distinguish it 
from a mixture of mesoscopic phases [18].

In the region of the existence of the Gibbs mixture, when $n_B$ changes 
from $n_p$ to $n_h$, the transition line is represented by a horisontal line
\be
T_c(n_B) =const \; \qquad (n_p\leq n_B\leq n_h)
\ee
connecting the points $T_h$ and $T_p$ defined in the relations (7). Along 
this line, the pressure
\be
p(T,n_B) \equiv P(T,\mu_B(T,n_B))
\ee
is given by a horisontal line connecting the points $P_h$ and $P_p$, 
according to equality (5),
\be
p(T_c,n_B) =const \qquad (n_p\leq n_B\leq n_h)\; .
\ee
Then the compression modulus
\be
\kp_T^{-1}(T,n_B) \equiv n_B\;\frac{\prt p}{\prt n_B}
\ee
remains zero on the transition line,
\be
\kp_T^{-1}(T_c,n_B) = 0 \qquad (n_p\leq n_B\leq n_h) \; .
\ee
Hence, the compressibility $\kp_T=\infty$ diverges exhibiting 
instability. Thus, the Gibbs mixture is, actually, {\it unstable}.

Constructing the pressure of a pure phase, one invokes phenomenological 
arguments of the mean-field type. The most popular is a kind of a 
quasiparticle description, when each sort of particles is characterized 
by an effective spectrum $\om_i(k)$, the interparticle interactions being 
included in the spectrum as mean-field parts. For example, one may employ 
a relativistic spectrum
$$
\om_i(k) =\sqrt{k^2 +m_i^2 +\Pi_i} 
$$
or a semi-relativistic one
$$
\om_i(k) =\sqrt{k^2 +m_i^2} + U_i \; ,
$$ 
where $m_i$ is a bare particle mass and $\Pi_i$ or $U_i$ are the real 
parts of self-energy playing the role of mean fields. The interparticle
interactions are taken sometimes in the excluded volume approximation, as 
it is done in the statistical bootstrap models [19] of hadrons. The 
interactions between hadrons can be modelled by various linear [4,12] or 
nonlinear [20] functions of density.

The quasiparticle picture is also used for the quark-gluon plasma phase 
[4,12,21,22] giving the description of pure gluodynamics and of 
zero-density chromodynamics in good agreement with numerical lattice 
calculations for the region of pure deconfined phase. However, the 
pure-phase models cannot correctly describe the whole process of 
deconfinement, always predicting a first-order phase transition, contrary 
to lattice simulations (see discussion in [4]).

\section{Extrapolating Equations of State}

Instead of invoking phenomenological arguments for describing the whole 
process of deconfinement, one sometimes resorts to the following method. 
One starts with considering the properties of a pure phase far from 
deconfinement, where the equation of state for this pure phase can be found 
with a reliable certainty, and then one extrapolates the found equation to 
the region close to deconfinement. As is evident, analyzing the properties 
of a sole pure phase, it is impossible to get a correct description of the 
whole phase transition, but only the region of stability can be determined 
in this way. Nevertheless, determining an instability point can give a 
reasonable estimate for that of deconfinement, and the extrapolated 
equation of state could provide an approximation for the region close to 
deconfinement. For example, examining the stability boundaries for nucleons 
inside nuclear matter yields [23] quite reasonable values for the 
instability temperature of 200 MeV and for the instability of 2 normal 
densities, which are close to the estimates for the deconfinement 
temperature and density [1--5].

The QCD pressure is known to be presentable as an asymptotic expansion in 
powers of the coupling parameter $g$ at high temperature and zero chemical 
potential, when the coupling parameter is small. This expansion is known 
[24,25] to the order of $O(g^6\ln g)$, 
\be 
P(g) \simeq \frac{8\pi^2}{45}\; T^4 \left ( a_0 + a_2g^2 +a_3 g^3 + a_4 g^4 +
a_4' g^4\ln g + a_5 g^5 \right ) \; ,
\ee
with the coefficients
$$
a_0 = 1 +\frac{21}{32}\; N_f \; , \qquad
a_2 = -0.09499\left ( 1 + \frac{5}{12}\; N_f\right ) \; ,
\qquad
a_3 = 0.12094 \left ( 1 + \frac{1}{6}\; N_f\right )^{3/2} \; ,
$$
$$
a_4 = 0.04331\left ( 1 + \frac{1}{6}\; N_f\right )\ln \left
( 1 + \frac{1}{6}\; N_f\right )  + 0.01733 - 0.00763 N_f - 0.00088 N_f^2-
$$
$$
- 0.01323 \left ( 1 + \frac{5}{12}\; N_f\right )
\left ( 1 -\; \frac{2}{33}\; N_f\right ) \; \ln\frac{\mu}{T} \; ,
\qquad
a_4' = 0.08662 \left ( 1 + \frac{1}{6}\; N_f\right ) \; ,
$$
$$
a_5 = - \left ( 1 + \frac{1}{6}\; N_f\right )^{1/2} \left (
0.12806 + 0.00717 N_f - 0.00027 N_f^2 \right ) + 
$$
$$
+0.02527 \left ( 1 + \frac{1}{6}\; N_f\right )^{3/2} \left
( 1 - \; \frac{2}{33}\; N_f\right ) \; \ln \frac{\mu}{T} \; ,
$$
where $N_f$ is the number of flavours. The dimensional regularization is 
used here, and the renormalization scale $\mu$ corresponds to the modified 
minimal substruction scheme $\overline{MS}$.

This expansion (16) is not convergent, but it is merely asymptotic, being 
valid only for $g\rightarrow 0$. Accepting the high-temperature dependence
$$
g^2 (T) \simeq 
\frac{24\pi^2}{(11N_c - 2N_f)\ln(T/\Lb)} \qquad (T\ra\infty) \; ,
$$
in which $N_c=3$ is the number of colours and $\Lb\approx 200$ MeV is the QCD 
scale parameter, one sees that the condition $g\ll 1$ corresponds to the 
very high temperatures $T>10^3$ MeV. This is why the form (16) does not 
agree with the lattice simulations for lower temperatures [26].

To extrapolate expression (16) to the region of finite $g$, Pad\'e 
approximants have been used [27,28]. The latter are often employed in the 
attempt of improving perturbative results of field theory [29]. However, 
the constructed Pad\'e approximants exhibit unnatural features, containing 
terms proportional to $g$ both in the numerator and denominator [28]. The 
most important is that the Pad\'e approximants do not converge, but some 
turn  out to develop unphysical poles [27,28]. At large $g$, Pad\'e 
approximants exhibit chaotic behaviour, since $P_{MN}\sim g^{M-N}$ as 
$g\ra\infty$, so that $P_{MN}\ra\pm\infty$ if $M>N$, $P_{MN}\ra const$ for 
$M=N$, and $P_{MN}\ra 0$ when $M<N$.

Another method of deriving expressions, valid at finite values of the 
coupling parameter, from asymptotic expansions having sense only in the 
vicinity of zero coupling parameter, is based on the {\it Self-Similar 
Approximation Theory} [30--39]. Below, we give a survey of the method 
allowing us to obtain the equation of state in QCD for finite values of the 
coupling parameter and for temperatures in a wide diapason [40]. The 
approach employs the {\it self-similar exponential approximants} [38], 
which, contrary to Pad\'e approximants, contain no poles and possess good 
convergence.

It is convenient to introduce the dimensionless function
\be
\ov P(g) \equiv \frac{P(g)}{P(0)} \; , \qquad
P(0) =\frac{8\pi^2}{45} \left ( 1 +
\frac{21}{32}\; N_f\right )\; T^4 \; ,
\ee
normalizing pressure $P(g)$ by the Stefan-Boltzmann limit $P(0)$. Then, 
expansion (16) reduces to the set of the approximants
\be
\ov P_k(g) = \sum_{n=0}^k \ov a_n g^n \; ,
\ee
where $k=0,1,2,3,4,5$ and the reduced coefficients are
$$
\ov a_0 = 1 \; , \qquad \ov a_1 = 0 \; , \qquad \ov a_4 =
\frac{a_4+a_4'\ln g}{a_0} \; , \qquad \ov a_n =\frac{a_n}{a_0} \qquad 
(n=2,3,5) \; .
$$

According to the idea of the {\it optimized perturbation theory} [30], the 
renormalization scale can be treated as a control function defined by the 
minimal difference condition. For the present case, we require that the 
approximation (18), where $\mu$ appears first, be equal to the precedent 
approximation. This leads  to the equations
\be
\ov P_4(g) =\ov P_3(g) \; , \qquad \ov a_4 = 0 \; .
\ee
Such minimal difference conditions are often employed in theoretical 
calculations [30,41--47]. The meaning of this condition has been explained 
in the frame of the self-similar approximation theory [30--39] as a kind of 
a fixed-point condition for an approximation cascade. Another type of 
fixed-point conditions is the minimal sensitivity condition [48--55] that 
is also often used in calculations. But the latter condition cannot be 
directly applied to the expansion (18). It is worth noting that the 
optimized perturbation theory [30] should not be confused with the 
variational minimization of free energy, common in statistical mechanics 
[56,57]. The optimized perturbation theory is a {\it systematic procedure} 
yielding a convergent {\it sequence of approximants}, while the variational 
minimization of free energy is a {\it one-step procedure} giving just a 
{\it single estimate}. In addition, the latter is valid solely for {\it 
free energy} or some other thermodynamic potential, since this variation is 
based on minimizing the right-hand side of the Gibbs-Bogolubov inequality, 
while the optimized perturbation theory can be developed for {\it any  
quantity} of interest [30].

Condition (19) results in the renormalization scale
\be
\mu = \gm T g^\nu = \mu(T,g)
\ee
as a function of temperature and the coupling parameter, where
$$
0.01323 \left ( 1 + \frac{5}{12}\; N_f \right )
\left ( 1 -\; \frac{2}{33}\; N_f \right ) \ln \gm =
$$
$$
= 0.04331 \left ( 1 + \frac{1}{6}\; N_f \right )
\ln \left ( 1 + \frac{1}{6}\; N_f \right ) + 0.01733 -
0.00763 N_f - 0.00088 N_f^2
$$
and
$$
\nu \equiv \frac{0.08662\left ( 1 +\frac{1}{6}\; N_f\right )}
{0.01323 \left ( 1 + \frac{5}{12}\; N_f \right )
\left ( 1 - \frac{2}{32}\; N_f \right )} \; .
$$
In particular, for $N_f =6$, one has $\gm=0.996964$ and $\nu=5.879155$. 
With the scale (20), the form (18) reduces to the expansion
\be
\ov P(g) \simeq 1 + \ov a_2 g^2 + \ov a_3 g^3 + \ov a_5 g^5 \; ,
\ee
valid for $g\ra 0$.

The extrapolation of the asymptotic expansion (21) to the region of finite 
$g$, by means of the self-similar exponential approximants [38,40], leads 
to the sequence
$$
{\ov P}_2^*(g) =\exp\left ( c_2 g^2\right ) \; ,
\qquad
{\ov P}_3^*(g) =\exp\left ( c_2 g^2\exp (c_3 g) \right ) \; ,
$$
\be
{\ov P}_5^*(g) =\exp\left ( c_2 g^2\exp (c_3 g\exp (c_5 g^2) ) \right ) \; ,
\ee
in which the coefficients
$$
c_2 =\frac{a_2}{a_0}\; \tau_2 \; , \qquad
c_3 =\frac{a_3}{a_2}\; \tau_3 \; , \qquad
c_5 =\frac{a_5}{a_3}\; \tau_5
$$
are connected with the control functions $\tau_i$. The latter are to be 
defined from fixed-point conditions or from the minimization of a cost 
functional.

The running QCD coupling $\al_s=\al_s(\mu)$ satisfies the 
renormalization-group equation
\be
\mu\; \frac{\prt\al_s}{\prt\mu} = \bt(\al_s) \; ,
\ee
which, because of the relation $g^2=4\pi\al_s$, defines the dependence of 
$g$ on $\mu$. The renormalization function $\bt(\al)$ is known for $\al\ra 0$ 
in the four-loop order [58] as the asymptotic expansion
\be
\bt(\al) \simeq b_2\al^2 + b_3\al^3 + b_4 \al^4 + b_5 \al^5 \; ,
\ee
with the coefficients
$$
b_2 = -\; \frac{1}{2\pi} \left ( 11 -\; \frac{2}{3}\; N_f\right ) \; ,
\qquad
b_3 = -\; \frac{4}{(4\pi)^2} \left ( 51 -\; \frac{19}{3}\; N_f\right ) \; ,
$$
$$
b_4 = -\; \frac{1}{(4\pi)^3} \left ( 2857 -\; \frac{5033}{9}\; N_f
+\frac{325}{27}\; N_f^2 \right ) \; ,
$$
$$
b_5 = -\; \frac{2}{(4\pi)^4} \left ( 29243 - 6946.3 N_f
+ 405.089 N_f^2 + 1.49931 N_f^3 \right ) \; ,
$$
In particular, for $N_f=6$, we have
$$
b_2 = -\frac{7}{2\pi} \; , \quad
b_3 = -\frac{52}{(4\pi)^2} \; , \quad
b_4 = \frac{65}{(4\pi)^3} \; , \quad
b_5 = -\frac{4944.50992}{(4\pi)^4} = -0.198282 \; .
$$
Generally, the signs of the coefficients $b_i$ depend on the number of 
flavours $N_f$ in the following way:
$$
b_2<0\; , \quad b_3 < 0 \; , \quad b_4 < 0 \; , \quad b_5 < 0 \qquad
(0\leq N_f \leq 5) \; ,
$$
$$
b_2<0\; , \quad b_3 < 0 \; , \quad b_4 > 0 \; , \quad b_5 < 0 \qquad
(6\leq N_f \leq 8) \; ,
$$
$$
b_2<0\; , \quad b_3 > 0 \; , \quad b_4 > 0 \; , \quad b_5 < 0 \qquad
(9\leq N_f \leq 16) \; ,
$$
$$
b_2>0\; , \quad b_3 > 0 \; , \quad b_4 > 0 \; , \quad b_5 < 0 \qquad
(17 \leq N_f \leq 40) \; ,
$$
$$
b_2>0\; , \quad b_3 > 0 \; , \quad b_4 < 0 \; , \quad b_5 < 0 \qquad
(41 \leq N_f <\infty ) \; .
$$
The qualitative change in the behaviour of $\bt(\al)$ happens at $N_f$ where 
the coefficient $b_2$ from negative becomes positive [59]. The coefficients 
$b_2$ and $b_3$ are renorm-scheme independent, but the higher coefficients 
$b_4$ and $b_5$ depend on the renorm-scheme employed  in their calculation. 
The expansion (24) is obtained [58] within the minimal substruction scheme.
But since the $\bt$- function does not depend explicitly on $\mu$, this 
function is the same in $\ov{MS}$ scheme.

Defining the reduced function
\be
\ov\bt(\al) \equiv \frac{\bt(\al)}{b_2\al^2} \; ,
\ee
we find from expansion (24)
\be
\ov\bt(\al) \simeq 1 +\ov b_3\al + \ov b_4\al^2 + \ov b_5\al^3 \; ,
\ee
with the reduced coefficients
$$
\ov b_n \equiv \frac{b_n}{b_2} \qquad (n=3,4,5) \; .
$$
The self-similar exponential approximants extrapolating Eq. (24) to finite 
$\al$ are
$$
\bt_3^*(\al ) = b_2 \al^2 \exp (d_3 \al) \; , \qquad
\bt_4^*(\al ) = b_2 \al^2 \exp (d_3 \al\exp(d_4 \al)) \; ,
$$
\be
\bt_5^*(\al ) = b_2 \al^2 \exp (d_3 \al\exp(d_4\al\exp(d_5\al))) \; ,
\ee
where the coefficients
$$
d_n\equiv \frac{b_n}{b_{n-1}}\; t_n \qquad (n=3,4,5)
$$
are expressed through the control functions $t_n$, which again have  to be 
defined either from fixed-point conditions or from the minimization of a 
cost functional.

Substituting the approximants (27) into the renorm-group equation (23), we 
solve the latter obtaining the corresponding approximations for 
$\al_s(\mu)$. As an initial condition, we may take the value
\be
\al_s(m_Z) = 0.1185 \; , \qquad m_Z=91.1882 \; {\rm GeV}
\ee
at the $Z^0$ boson mass [60].

The  renorm-group equation (23), with the right-hand side defined by one of 
the forms (27), gives $\al_s=\al_s(\mu)$. The relation $g^2=4\pi\al_s$, 
together with the scale (20), leads to the equation
\be
g^2 = 4\pi\al_s(\mu(T,g))
\ee
determining the function $g=g(T)$. Substituting the latter in the 
approximants (22) results in the reduced pressure
\be
\ov p_k(T) \equiv {\ov P}_k^*(g,T)
\ee
as a function of temperature.

Calculations show [40] that the behaviour of the reduced pressure $\ov 
p(T)$ is in reasonable agreement with lattice simulations [4]. At the 
temperature $T_c\approx 200$ MeV, the pressure sharply drops down, when 
decreasing $T$, which can be interpreted as confinement. However, the 
details of the confinement-deconfinement process cannot be accurately 
described by such an extrapolation approach based on the consideration of 
the quark-gluon plasma only. In addition, the self-similar exponential 
approximants (22) or (27) provide an accurate extrapolation to the region 
of finite $g$ or $\al_s$, where these parameters are of order one. In the 
vicinity of confinement, when $\al_s$ fastly grows becoming much more than 
one, any extrapolation procedure would be quantitatively unreliable.

\section{Effective Coupling under Confinement}

The running coupling $\al_s(\mu)$ as a function of the scale $\mu$ is 
experimentally studied only for $\mu\geq 2$ GeV, where $\al_s<0.4$ [60,61].
In the low-momentum region, the behaviour of the effective coupling is 
poorly known not because of the limited knowledge of higher order effects, 
but because of an essentially different physical phenomenon that enters the 
game, the one that is referred to as confinement.

For large $\mu$, perturbation theory gives
\be
\al_s(\mu) \simeq \frac{2\pi}{\bt_0\ln(\mu/\Lb)} \qquad (\mu\ra\infty) \; ,
\ee
where $\Lb\approx 200$ MeV is the QCD scale parameter and $\bt_0\equiv 
-2\pi b_2=11-\frac{2}{3}N_f$. The form (31) is valid if $\al_s\ll 1$, that 
is when 
\be
\mu \gg \Lb\exp\left ( \frac{2\pi}{\bt_0} \right ) > \Lb \; .
\ee
This implies, as far as $\Lb\approx T_c\approx 200$ MeV, that exression (31) 
is applicable only for $\mu\gg T_c$. If, nevertheless, one formally 
considers (31) at lower $\mu$, then the coupling (31) diverges at $\mu=\Lb$.
There exist arguments [62] that $\al_s(\mu)$ is finite at all $\mu$, and 
satisfies the sum rule
$$
\frac{1}{\pi} \int_0^{2{\rm GeV}}\al_s(k)\; dk \approx 0.38\; {\rm GeV} \; .
$$
There are several models [62,63] constructing $\al_s(\mu)$ for arbitrary $\mu$.

The simplest way of a phenomenological construction of finite coupling 
could be by means of the {\it pole-removal trick}. The idea of the latter 
is as follows. Suppose that a function $f(x)$ has a pole at $x_0$, which 
implies that in the vicinity of the pole the function can be presented as 
the sum
\be
f(x) \simeq f_{reg}(x) + f_{sin}(x) \qquad (x\ra x_0)
\ee
of a regular and singular parts. Let us define a {\it regularized function}
\be
\tilde f(x) \equiv f(x) - f_{sin}(x)
\ee
as the function $f(x)$, with the removed singular part.

In applying this trivial trick to the running coupling, we may {\it 
postulate} that the latter is defined by extrapolating the perturbative 
approximation (31) to all $\mu$ by means of the regularization procedure (34).
The perturbative expression (31) can be identically rewritten as
$$
\al_s(\mu) \simeq \frac{2\pi n}{\bt_0\ln(\mu/\Lb)^n} \; ,
$$
with any positive $n>0$. Taking into account the asymptotic equality
$$
\frac{1}{\ln x}\simeq \frac{1}{2}\; - \; \frac{1}{1-x} \qquad (x\ra 1) \; ,
$$
we immediately obtain the {\it regularized coupling} 
\be
\tilde\al_s(\mu) =\frac{2\pi n}{\bt_0} \left [ 
\frac{1}{\ln(\mu/\Lb)^n} + \frac{1}{1-(\mu/\Lb)^n} \right ] \; ,
\ee
which is finite for any $\mu$, including $\mu=\Lb$ and $\mu=0$, where
\be
\tilde\al_s(0) =\frac{2\pi n}{\bt_0} \; , \qquad \tilde\al_s(\Lb) =
\frac{\pi n}{\bt_0} \; .
\ee
Similarly, one can construct regularized couplings from perturbative 
expressions of higher orders. But let us note that, as follows from Eq. 
(35), the pole-removal trick does not define the regularized functions in a 
unique way.

Another way of regularization, leading to the same result, would be by 
means of {\it analytical continuation}. Consider again a function $f(x)$, 
with $x\geq 0$, having one or several poles on the positive semiaxis. 
Assume that $f(-x)$ has no poles. Define the analytic continuation $\tilde 
f(z)$ to the complex plane, except the cut along the negative semiaxis, so that
\be
\tilde f(-x\pm i0) = f(-x \pm i0) \qquad (x\geq 0) \; .
\ee
In the region of analyticity of $\tilde f(z)$, the spectral representation
\be
\tilde f(z) =\frac{1}{\pi} \; \int_0^\infty 
\frac{J(x)}{x+z} \; dx
\ee
is valid, from where the spectral function is
\be
J(x) =\frac{i}{2} \left [ \tilde f(-x +i0) -\tilde f(-x -i0)\right ] \; .
\ee
In the latter, condition (37) is to be used.

Applying the analytic continuation method to the perturbative coupling 
(31), we take into account that $\ln(-x\pm i0)=\ln|x|\pm i\pi$ and
$$
\frac{1}{\ln(-x\pm i0)} = \frac{\ln|x|\mp i\pi}{\ln^2|x| +\pi^2} \; .
$$
Then the corresponding spectral function is
\be
J(x) =\frac{2\pi^2 n}{\bt_0(\ln^2|x| +\pi^2)} \; , \qquad x\equiv \left (
\frac{\mu}{\Lb}\right )^n \; .
\ee
The spectral representation (38) gives the same regularized coupling (35). 
This way of regularization was employed in Refs. [64,65] for the case $n=2$.

The weakest point in any regularization procedure, based on the 
perturbative expression (31), is that the latter is valid only under 
condition (32), hence a  regularized function has sense solely for 
$|\mu/\Lb|\gg 1$. And there is no any reason of extrapolating $\al_s(\mu)$ 
to the region $\mu/\Lb\leq 1$, where confinement is expected.

\section{Clustering Quark-Hadron Matter}

Any attempt of treating deconfinement from the point of view of pure 
thermodynamic phases contains the following {\it principal contradiction}. 
Hadrons are believed to present {\it bound states} of quarks and gluons, 
while quark-gluon plasma represents their {\it unbound states}. This 
implies that the system of quarks and gluons possesses, in general, both a 
discrete spectrum corresponding to bound states and a continuous spectrum 
associated with unbound states. When a many-body system possesses an 
energy spectrum $E_n$, then the distribution of particles over the energy 
levels is described by the Gibbs probability $p_n\sim e^{-\bt E_n}$, where 
$\bt T\equiv 1$. If the energy spectrum contains both discrete as well as 
continuous parts, at each moment of time there exists a probability for 
particles to form bound states or to pertain to unbound states. In the 
standard quantum picture, bound and unbound states do coexist, with the 
related probability weights. Hence, hadrons must coexist with quark-gluon 
plasma. Let us stress that this is a direct logical conclusion immediately 
resulting from the treatment of hadrons as describing bound quark-gluon 
states. If a quantum system possesses different parts of spectra, all of 
them are to be taken into account by calculating the corresponding 
probability weights. It is not correct to separate the whole spectrum onto 
particular sections, prohibiting the existence of some of its parts. It is 
also incorrect to identify separate sections of the quantum spectrum with 
different thermodynamic phases. {\it Hadron states and plasma states are 
quantum states but not thermodynamic phases}.

This situation is similar to that of electron-ion plasma. Electrons and 
ions can form bound states, i.e. neutral atoms, or unbound states of 
electrons and charged ions. In the system of electrons and ions, under 
given conditions, there is a fraction of neutral atoms and a portion of 
separated ions and electrons. Changing conditions varies the fractional 
concentrations of neutral and ionized atoms. Ionization in the system of 
predominantly neutral atoms is the direct analog of deconfinement in 
predominantly hadronic matter. Ionization as well as deconfinement can 
occur, depending on circumstances, either as a sharp transition or as a 
gradual crossover.

The description of statistical properties of a quantum many-body system 
possessing several qualitatively different quantum states, such as bound 
and unbound, is not a trivial task. For describing such systems, {\it 
Theory of Clustering Matter} has been elaborated [4,12]. The approach is 
based on three pivotal concepts: {\it Cluster Representation}, {\it Statistical 
Correctness}, and {\it Potential Scaling}.

The idea of the quasiparticle {\it cluster representation} goes back to the 
authors who analyzed the abundances of chemical elements on Earth by 
treating each element as a quasiparticle characterized by the corresponding 
atomic weight and the binding energy, with the related chemical potentials
taking into account the allowed interparticle reactions. Such approaches 
are reviewed in Refs. [66--68]. The same idea was applied to considering 
nuclear multifragmentation [69]. A more accurate mathematical formulation 
for the problem of constructing the quasiparticle representation for 
composite particles was initiated by Weinberg [70--72]. Such a 
representation could be unambiguously defined provided that a 
transformation from the state space of elementary particles to that of the 
system containing composite particles, together with unbound elementary 
particles, would be given [73--75]. For this purpose, different Boson 
realizations of Lie algebras [76] were employed [77--81]. The most general 
approach, based on the Tani transformation [82], has been developed by 
Girardeau [83--86] who coined the term {Fock-Tani representation}. This was 
applied to various systems containing bound clusters, including the 
quark-hadron matter [87].

The basic point in the quasiparticle cluster picture is as follows. 
Consider a many-body system, with the total space of quantum states being a 
Fock space ${\cal F}$. Let the algebra of observables, ${\cal A}$, be 
defined on ${\cal F}$. Assume that the particles of the system can form 
several types of bound states, e.g. corresponding to different hadron clusters. 
Enumerate all admissible types of bound states by the index $i=2,3,\ldots$, 
reserving the index $i=1$ to unbound states. Each kind of bound clusters 
can be individualized by a set of characteristic parameters, such as the 
compositeness number $z_i$ showing the number of elementary particles bound 
into a cluster, effective mass of the cluster $m_i$, and a set of  quantum 
numbers like spin, isospin, colour, baryon number, strangeness, and so on. 
And let us treat each type of bound clusters as a separate sort of 
particles, with the associated Fock space ${\cal F}_i$, called the ideal 
cluster space. The direct product
\be
\tilde{\cal F} \equiv \otimes_i {\cal F}_i \qquad ({\cal F}_1={\cal F})
\ee
composes the {\it total cluster space}. The formal relation between the 
Fock space of elementary-particle states and the cluster space (41) can be 
presented by means of a unitary transformation $\hat U$, such that
\be
{\cal F}=\hat U \tilde{\cal F} \; , \qquad 
\tilde{\cal F} =\hat U^+{\cal F} \; . 
\ee
Then the {\it cluster algebra of observables} is defined as 
\be
\tilde{\cal A} \equiv \hat U^+{\cal A}\hat U \; .
\ee
With these definitions, all matrix elements
of the algebra ${\cal A}$ in ${\cal F}$ are the same as those of $\tilde{\cal 
A}$ in $\tilde{\cal F}$, since $\tilde{\cal F}\tilde{\cal A}\tilde{\cal F} = 
{\cal F}{\cal A}{\cal F}$. Since the representations of $\tilde{\cal A}$ in 
$\tilde{\cal F}$ and ${\cal A}$ in ${\cal F}$ are isomorphic, all observables 
quantities are the same in the standard picture of elementary particles and in 
the quasiparticle picture of a clustering system.

Let us now delineate the mathematical structure of the Tani transformation. 
Let the field operators of elementary particles, say of quarks, be $q(x)$ 
defined on the Fock space ${\cal F}$, with $x$ being a set of spatial variables.
Suppose $\varphi_i(x_1,x_2,\ldots,x_i)$ is a Schr\"odinger wave function 
describing a bound state of $i$ elementary particles. The field operator of 
this bound state can be presented as
$$
\Psi_i(x) \equiv \int\; \varphi_i(x_1-x,x_2-x,\ldots,x_i-x)\; q(x_1) \;
q(x_2)\ldots q(x_i) \; dx_1dx_2\ldots dx_i \; .
$$
The image of the bound state in the ideal cluster space ${\cal F}_i$ is given 
by a 
cluster with the field operator $\psi_i(x)$. By definition, $\psi_1(x)\equiv
q(x)$. The Tani transformation is described by the unitary operator
\be
\hat U =\exp\left (\frac{\pi}{2}\;\hat F\right ) \; , \qquad
\hat F =\sum_i\int\left [ \psi_i^\dgr(x)\Psi_i(x) - 
\Psi^\dgr_i(x)\psi_i(x)\right ] \; dx \; .
\ee

In the cluster representation, constructed on the cluster space (41), one 
defines the statistical state $<\tilde{\cal A}>$ for the algebra of 
observables (43). The density of the $i$- type clusters is
$$
\rho_i =\frac{1}{V}\; <\hat N_i> \; , \qquad
\hat N_i =\int\psi^\dgr_i(x)\psi_i(x)\; dx \; .
$$
The probability for this type of clusters to be formed can be characterized 
by the weight
\be
w_i= z_i\; \frac{\rho_i}{\rho} \qquad \left ( \rho\equiv \sum_i z_i\rho_i
\right ) \; ,
\ee
which may be called the {\it cluster probability}. The weight (45) 
satisfies the standard properties of probability, being nonnegative, 
$0\leq w_i\leq 1$, and normalized, $\sum_i w_i=1$.

The direct calculation of the {\it cluster Hamiltonian}
\be
\tilde H = \hat U^+ H\hat U
\ee
is a rather complicated problem. Moreover, the actual form of the 
Hamiltonian (46) is written as an infinite series. Because of this, one 
usually simplifies the procedure by assuming an effective Hamiltonian 
$H_{eff}$, whose construction involves physical reasoning. The latter 
Hamiltonian is often written with an explicit dependence on thermodynamic 
parameters, such as the cluster densities $\rho_i$ and temperature $T$, so 
that $H_{eff}=H_{eff}(\{\rho_i\},T)$. At this point the principle of {\it 
statistical correctness} [4,12] comes into play saying that the general 
form of the cluster Hamiltonian (46) has to be as
\be
\tilde H =H_{eff} + CV \; ,
\ee
where $C$ is a nonoperator term such that makes the Hamiltonian(47) 
statistically correct, which implies the validity of the equations
\be
<\frac{\prt\tilde H}{\prt\rho_i}> \; = 0 \; , \qquad
<\frac{\prt\tilde H}{\prt T}> \; = 0 \; .
\ee
From Eqs. (47) and (48) one gets the equations
\be
\frac{\prt C}{\prt\rho_i} = -\; \frac{1}{V}\; 
<\frac{\prt H_{eff}}{\prt\rho_i} > \; , \qquad
\frac{\prt C}{\prt T} = -\; \frac{1}{V}\; 
<\frac{\prt H_{eff}}{\prt T}> \; ,
\ee
defining $C=C(\{\rho_i\},T)$. These conditions garantee the validity of the 
thermodynamic relations
$$
P = -\; \frac{\prt\Om}{\prt V} = - \; \frac{\Om}{V} \; , \qquad
\ep = T\; \frac{\prt P}{\prt T}\; - P  +\mu_Bn_B +\mu_S n_S =
\frac{1}{V}\; <\hat E>\; ,
$$
$$
s=\frac{\prt P}{\prt T} =\frac{1}{T}\left (\ep + P - \mu_B n_B - n_S n_S
\right ) \; , \quad
n_B =\frac{\prt P}{\prt\mu_B} = \sum_i B_i\rho_i \; , \quad
n_S =\frac{\prt P}{\prt\mu_S} = \sum_i S_i\rho_i \; ,
$$
in which $\ep$ and $s$ are the energy and entropy densities, and
$$
\hat E =\tilde H + \sum_i \mu_i\hat N_i \; , \qquad
\rho_i =\frac{\prt P}{\prt\mu_i} =\frac{1}{V}\; <\hat N_i> \; .
$$

The cluster Hamiltonian (47) contains the terms describing effective 
interactions between different clusters. For defining the corresponding 
interaction potentials, the principle of {\it potential scaling} has been 
formulated [4,12]. According to the latter, the interaction potentials from 
the same class of universality are connected by the scaling relation
\be
\frac{\Phi_{ij}({\bf r})}{z_iz_j} = \frac{\Phi_{ab}({\bf r})}{z_az_b} \; .
\ee
This allows the definition of all qualitatively similar interaction 
potentials through one known potential. Another form of scaling (50) could be
$$
\frac{\Phi_{ij}({\bf r})}{m_im_j} =
\frac{\Phi_{ab}({\bf r})}{m_a m_b} \; ,
$$
provided that $m_i\sim z_i$.

The theory of clustering matter has been applied to {\it clustering 
quark-hadron matter} [4,12]. The appearance of multiquark clusters in 
nuclear matter is explained. The possibility of the dibaryon Bose 
condensation is advanced. Provisions for nuclear-matter lasers are 
estimated [88]. Thermodynamic characteristics for the SU(3) gluodynamics and 
zero-baryon-density chromodynamics are in good quantitative 
agreement with lattice simulations, displaying a first-order transition for 
pure gluodynamics and a crossover for chromodynamics. Deconfinement at 
finite baryon density, at conditions typical of heavy-ion collisions, is 
predicted to be a gradual crossover.

\section{Discussion}

The model of clustering quark-hadron matter [4,12] provides, to our mind, 
the most realistic approach to describing deconfinement at finite baryon
density. Deconfinement is found to be a gradual crossover, but not a sharp 
transition. However, it would be yet too premature to state that all 
details of the deconfinement process are well understood. We do not imply 
here some technicalities that could always be varied in the frame of the 
same general approach to describing the clustering quark-hadron matter. For 
instance, one can take different interaction potentials or accept different 
cluster spectra $\om_i(k)$. Such technical variations do not change the 
general qualitative picture. But there are more principal questions that 
have not yet been properly addressed:

\vskip 1mm

(i) The clustering quark-hadron matter has been treated in the mean-field
approximation [4,12]. This seems to be reasonable especially because 
deconfinement is not a second-order phase transition but rather a 
crossover. The state where unbound quarks and gluons coexist with hadron 
clusters is shown to be thermodynamically stable with respect to the 
thermal and mechanical stability and with respect to the minimality of the 
thermodynamic potential as compared to those of pure phases corresponding 
either to quark-gluon plasma or to pure hadron states. However, the dynamic 
stability of the clustering system, which requires the positiveness of the 
collective-excitation spectrum, has not been checked. The latter would be 
suitable to consider as far as in the mean-field approximation the 
conditions of thermodynamic and dynamic stability do not coincide.

\vskip 1mm

(ii) The finite-size effects in a clustering system have not been analyzed, 
while this would be useful keeping in mind that heavy ions colliding in 
realistic experiments are always finite. The finiteness of a system not only 
leads to quantitative corrections, as compared to an infinite matter, but 
may sometimes cause the existence of thermodynamic quasi-phases [89], not 
existing in thermodynamic limit.

\vskip 1mm

(iii) It is possible that small static bubbles of quark-gluon plasma inside 
the predominantly hadron matter could arise, and vice versa, static hadron 
bubbles inside  quark-gluon plasma could exist [90]. Also, static droplets of
strange matter, the so-called strangelets could be formed. All such 
possibilities should be considered in the framework of the clustering matter.

\vskip 1mm

(iv) Mesoscopic heterophase fluctuations may emerge even in a globally 
equilibrium system [18]. Such dynamically fluctuating germs of quark-gluon 
plasma or hadron droplets are principally different from static bubbles 
[90] and require a different theoretical approach [18].

\vskip 1mm

(v) Finally, to be closer to collision experiments, one should consider a 
nonstationary picture, analyzing all different possibilities mentioned 
above. In the nonequilibrium case, several scenarios of deconfinement 
could be feasible. Then the problem of pattern selection would arise, 
requiring the necessity of defining the probabilistic weights for the 
admissible deconfinement scenarios.

\vskip 2mm

Summarizing the main material of this review, we come to the following 
conclusions:

\begin{enumerate}

\item
For the correct description of deconfinement, it is necessary to employ 
the approach based on the clustering quark-hadron matter.

\item
The process of deconfinement has not yet been completely understood.

\item
One must be very cautious in trying to interpret observed experimental data 
as attributed to deconfinement.

\end{enumerate}

\end{document}